\documentclass[notitlepage,aps,prb,twocolumn,superscriptaddress]{revtex4-1}

\usepackage{graphicx}  
\usepackage{subfigure}
\usepackage{dcolumn}   
\usepackage{bm}        
\usepackage{amssymb}   
\usepackage{comment}
\usepackage{hyperref}
\usepackage{color}
\usepackage{amsmath}
\usepackage{mathrsfs}
\usepackage{mathcomp}
\usepackage{textcomp}
\usepackage{dsfont}
\usepackage{esint}
\usepackage{braket}
\usepackage{textgreek}
\usepackage{lipsum}

\begin{document}

\title{Electronic scattering, focusing, and resonance by a spherical barrier in Weyl semimetals}

\author{Ming Lu}
\affiliation{International Center for Quantum Materials and School of Physics, Peking University, Beijing 100871, China}
\affiliation{Collaborative Innovation Center of Quantum Matter, Beijing 100871, China}
\author{Xiao-Xiao Zhang}
\affiliation{Department of Applied Physics, The University of Tokyo, Tokyo 113-8656, Japan}


\newcommand\dd{\mathrm{d}}
\newcommand\ii{\mathrm{i}}
\newcommand\ee{\mathrm{e}}
\newcommand\zz{\mathtt{z}}
\makeatletter
\let\newtitle\@title
\let\newauthor\@author
\def\ExtendSymbol#1#2#3#4#5{\ext@arrow 0099{\arrowfill@#1#2#3}{#4}{#5}}
\newcommand\LongEqual[2][]{\ExtendSymbol{=}{=}{=}{#1}{#2}}
\newcommand\LongArrow[2][]{\ExtendSymbol{-}{-}{\rightarrow}{#1}{#2}}
\newcommand{\cev}[1]{\reflectbox{\ensuremath{\vec{\reflectbox{\ensuremath{#1}}}}}}
\newcommand{\red}[1]{\textcolor{red}{#1}} 
\newcommand{\mycomment}[1]{} 
\makeatother

\begin{abstract}
We solve the Weyl electron scattered by a spherical step potential barrier. Tuning the incident energy and the potential radius, one can enter both quasiclassical and quantum regimes. Transport features related to far-field currents and integrated cross sections are studied to reveal the preferred forward scattering. In the quasiclassical regime, a strong focusing effect along the incident spherical axis is found in addition to optical caustic patterns. In the quantum regime, at energies of successive angular momentum resonances, a polar aggregation of electron density is found inside the potential. The findings will be useful in transport studies and electronic lens applications in Weyl systems.
\end{abstract}
\maketitle



\section{Introduction}%
The Weyl fermion\cite{Weyl1929} is a three-dimensional (3D) analogue of the two-dimensional (2D) Dirac physics\cite{graphene,*DiracFermion1,*DiracFermion2}. Different from the real-space monopoles realized as magnetic topological defects\cite{XXZ:resistivity,*XXZ:monopole}, a Weyl point can be identified as a momentum-space monopole\cite{Volovik,EEMF0}. Following the theoretical predictions through the breaking of time-reversal or inversion symmetry\cite{Volovik,Weyl2007,Weyl2011}, it was first realized in a family of nonmagnetic and noncentrosymmetric transition metal monoarsenides/monophosphides 
\cite{predict2,*predict1,*TaAS1,*TaAS2}. These advances also sparked other Weyl-like systems in terms of photonic\cite{WeylPhotonic1}, magnonic\cite{,
WeylMagnon5,*WeylMagnon1,*WeylMagnon2,*WeylMagnon3,*WeylMagnon4} and driven electronic bands\cite{\mycomment{Xing,*}XXZ:Floquet,*ZhongWang,*Rubio,*WeylMagnon6\mycomment{,*DanielLoss,*Awadhesh,*Oka}}. Based on the solid-state realization as a Weyl semimetal, many new phenomena are under intensive investigations both theoretically and experimentally\cite{WeylReviewQi,*WeylReviewBurkov1,*WeylReviewMagneto,*WeylReviewTransport,
WeylReviewBurkov2,*WeylReviewJia,*WeylReviewYan,*WeylReviewBurkov3,*WeylDiracReview}.

Under certain circumstances, the Weyl point as a topologically protected object is robust against weak disorders\cite{WeylImpurity1,*WeylImpurity2,*WeylImpurity3,*WeylImpurity4,*XXZ:Luttinger} and weak short-range\cite{WeylCDW1,*Nandkishore,*Juricic2017} or long-range\cite{Sekine\mycomment{,*Aji2014},*XXZ:meanfield} electron correlations. It warrants theoretical studies taking the 3D nondegenerate massless particle as a starting point since one would expect interesting physical effects arisen from such a new quasiparticle. In this spirit, here we would like to address a quantum mechanical problem of the simplest setting, i.e., a Weyl electron scattered by a spherically symmetric step potential barrier. On one hand, this scenario facilitates the exact solution that helps to unambiguously inspect the system. On the other hand, it is relevant to various problems of practical interests, especially as the advancing fabrication techniques of Weyl semimetals are paving the way for more controllable scientific explorations and possible device applications. Some related 
 considerations first arose in the graphene system\cite{2DCaustics,2DMie1,2DScatterCircular,2DScatterImpurity,2DLens,Review2DOptics} and relevant experimental observations have been reported\cite{2DMie2,whispering-gallery,grapheneFocusing}.

In reality, the potential profile varies mainly within a characteristic length scale $\lambda_V$. The step potential model is justified when the Fermi wavelength $\lambda_F$ and the lattice constant $d$ are respectively much larger and smaller than $\lambda_V$, where the latter is to suppress the internode scatterings. This setting can model the scattering by a single impurity or a cluster of impurities of the scalar (chemical potential) type and also the scattering by an electrostatically gated region in the material. As to the former, while a small-size single impurity is ubiquitous, a relevant case with clusters of various absorbates\cite{2DCluster} formed might be thin-film Weyl materials, which will be favored in building electronic and spintronic devices. The latter is possible in heterogeneously doped junction structures, in which a gate voltage much less affects the electrode contact doped outer $n$-region and produces an inner $p$-region of higher potential. This was used to build quantum dots in Dirac semimetals\cite{XXZ:nanowire2}, but for the nonce, 
probably only approximately spherical gating is technically feasible. In addition, realizations in other bosonic artifical Weyl systems like photonic crystals\cite{WeylPhotonic1} are possible as well. In general, this model serves as a working approximation towards more realistic and complicated situations. 

In this regard, the present study will not only help to capture certain effects due to the impurity scattering from a simplistic microscopic model, but also bring up the possibility of electronic lens applications featured by the negative refractive index based on Weyl materials. Indeed, after constructing the spherical eigenstates of the Weyl Hamiltonian and solving the potential scattering in Sec.~\ref{Sec:solution}, we will use the exact solution to discuss related aspects in Sec.~\ref{Sec:results}. A particular convenience is that we can vary the incident energy and potential radius to enter both quasiclassical and more quantum regimes. We successively discuss transport features related to the far-field current density and integrated cross sections, a new focusing along the high-symmetry axis of incidence and caustic effects, and a unique resonant polar aggregation of electron density inside the potential barrier. As a whole, the 3D Weyl semimetal shows distinguishably new features compared with the graphene system and hence is well worth more investigations of the scientific implication and the advantage in device applications.

%

\section{Spherical eigenstates and scattering of a Weyl electron}\label{Sec:solution}

The simplest Weyl electron is given by the Hamiltonian\cite{Volovik,Weyl2007,Weyl2011} $H_0=v\vec{\sigma}\cdot \vec{p}$ where $v$, $\vec{\sigma}$, and $\vec{p}=-\ii\hbar\nabla$ are the Fermi velocity, the Pauli matrices, and the momentum operator, respectively. To study the plane wave scattering of a Weyl electron by a spherical step potential $V(r)= V_0\theta(a-r)$, which is described by
\begin{equation}
H=v\vec{\sigma}\cdot \vec{p}+V(r),
\end{equation}
we should make use of the spherical rotation symmetry. It is essential to notice that this Hamiltonian conserves the `total angular momentum' $\vec{J}\equiv \vec{L}+\hbar\vec{\sigma}/2$ although $\vec{\sigma}$ can refer to either the real electron spin or some pseudospin degree of freedom. 
The `orbital angular momentum' $\vec{L}$ is not conserved, although useful as shown below. 

In order to obtain the simultaneous eigenstates of the complete set of commuting observables (CSCO) $\{H, J^2, J_z\}$, we detour to solve for $H^2$ in the first place. When $r>a$, we have $H^2=v^2\hat{p}^2 I_2$, where $\hat{p}^2=-\frac{1}{\hbar^2r^2}\frac{\partial}{\partial r}(r^2\frac{\partial}{\partial r})+\frac{1}{r^2}L^2$ and $I_2$ is the $2\times2$ identity matrix.
Separating variables, we can solve the simultaneous eigenstates of another spinless CSCO $\{H^2,L^2,L_z\}$. The general solution of the radial equation is a linear combination of the spherical Bessel functions of the first and second kind, $j_l(kr)$ and $n_l(kr)$, respectively. Since we expect the scattered or reflected wavefunction to hold the asymptotic form\cite{Sakurai} $\psi^\mathrm{ref}(r\to\infty)\sim f(\theta,\phi)\frac{\ee^{\ii kr}}{r}$
with scattering amplitude $f(\theta,\phi)$, polar angle $\theta$ and azimuthal angle $\phi$ with respect to the incident direction,
we can use the spherical Hankel function of the first kind, $h_l^{(1)}=j_l+\ii n_l$, as the radial wavefunction. Thus, the simultaneous eigenstates of $\{H^2,L^2,L_z\}$ take the form $\varphi_{k,l,m}=h_l^{(1)}(kr)Y_l^m(\theta,\phi) \times \text{arbitrary 2-component spinor}$, where $Y_l^m(\theta,\phi)$ is the conventional spherical harmonics. Based on this, we can further construct two simultaneous eigenstates of $\{H^2,J^2,J_z\}$, i.e., 
$\varphi_{k,j=l\pm\frac{1}{2},m_j}^{(\pm)} = h_{l}^{(1)}(kr)\mathcal{Y}_{jm_j}^{(\pm)}$ by defining the spin-\textonehalf \space angular wavefunctions
$\mathcal{Y}_{j=l\pm\frac{1}{2},m_j}^{(\pm)} = \frac{1}{\sqrt{2l+1}}\left(
\pm\sqrt{l\pm m_j+\frac{1}{2}}Y_{l}^{m_j-\frac{1}{2}},
\sqrt{l\mp m_j+\frac{1}{2}}Y_{l}^{m_j+\frac{1}{2}}
\right)^T$.
Note that $j$ is half-integer, $\braket{\varphi_{kjm_j}^{(+)}|\varphi_{kjm_j}^{(-)}}=0$ and the eigenvalues of $\varphi_{kjm_j}^{(\pm)}$ are the same $(\hbar v k)^2,j(j+1),m_j$ for the three operators.
This means that $\varphi_{kjm_j}^{(\pm)}$ are \textit{degenerate}. In fact, only when the parity operator $\mathcal{P}$ is added, $\{H^2, J^2, J_z, \mathcal{P}\}$ forms a CSCO since $\varphi^{(\pm)}$ is even/odd when $j+\frac{1}{2}$ is odd. 

Now we claim that $\psi_{kjm_j}^{(\pm)}=(H+\hbar v k)\varphi_{kjm_j}^{(\pm)}$ are the longed-for simultaneous eigenstates of CSCO $\{H, J^2, J_z\}$ with eigenvalues $\hbar vk$, $j(j+1)$ and $m_j$, respectively, which can be verified with 
the commutation relations.
On the other hand, one observes that $H\varphi_{kjm_j}^{(\pm)}=\mp\ii\hbar vk \varphi_{kjm_j}^{(\mp)}$. Consequently, we have $\psi_{kjm_j}^{(-)}=\hbar v k(\varphi_{kjm_j}^{(-)}+\ii \varphi_{kjm_j}^{(+)})$ while $\psi_{kjm_j}^{(-)}=\ii\psi_{kjm_j}^{(+)}$.
Therefore, $\psi_{kjm_j}^{(\pm)}$ are the \textit{same} state up to a global phase, which is certain since $\{H, J^2, J_z\}$ already forms a CSCO. We will simply adopt $\psi_{kjm_j}^{(-)}/(\hbar vk)$ henceforth as the eigenstate $\psi_{kjm_j}$ and when $r<a$, we need only change $k$ to $k'=k-V_0/(\hbar v)$ and $h_l^{(1)}$ to $j_l$ because $n_l$ diverges at the origin.
In summary, the spherical eigenstates of a Weyl Hamiltonian under a step potential are
\begin{align}
\psi_{kjm_j}^>(r>a) &= h_{j+\frac{1}{2}}^{(1)}(kr)\mathcal{Y}_{jm_j}^{(-)}+\ii  h_{j-\frac{1}{2}}^{(1)}(kr)\mathcal{Y}_{jm_j}^{(+)}\label{eq:sphere_eigen_in}
\\
\psi_{kjm_j}^<(r<a) &= j_{j+\frac{1}{2}}(k'r)\mathcal{Y}_{jm_j}^{(-)}+\ii j_{j-\frac{1}{2}}(k'r)\mathcal{Y}_{jm_j}^{(+)},\label{eq:sphere_eigen_out}
\end{align}
which are valid for both positive and negative arguments. This approach is useful when studying Dirac-like systems with symmetries. For instance, we find it possible to adapt and solve the scattering by a magnetic monopole that was previously obtained in a different manner\cite{CNYang2,*WeylMonopoleScattering}.

\begin{figure*}[htbp]
	\centering		
	\includegraphics[scale=0.65]{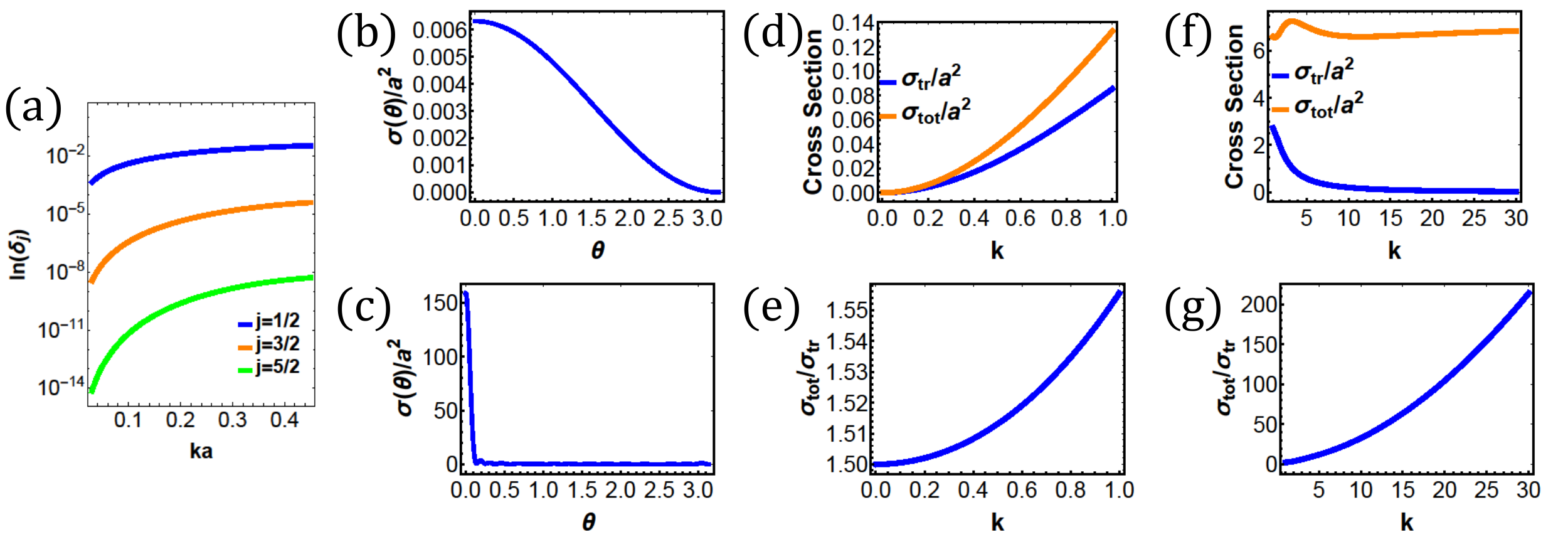}
		\caption{(a) Phase shifts of different partial waves at small incident electron energy with $a=1,V_0=1$\mycomment{$\frac{aV_0}{\hbar v}=1$}. (b,c): Differential cross section or far-field radial current density when the barrier radius is small (b) $a=0.5$ or large (c) $a=50$ with $k=0.5$ and $V_0=1$. (d,e): Scattering cross sections for small radius and low energy with $a=0.5, V_0=1$. (f,g): Scattering cross sections for larger radius and higher energy with $a=5,V_0=1$.}\label{fig:sigma}
\end{figure*}
The plane wave of a Weyl fermion adopts a spin-momentum locked form. Supposing that the incident motion is along the positive $z$-axis, we have $\psi^\mathrm{inc} = \ee^{\ii kz} \ket{z\uparrow}$ with $\ket{z\uparrow}=(1,0)^T$. With the asymptotic form and the finite value at the origin, we should expand it using \eqref{eq:sphere_eigen_out} with $k'$ replaced by $k$
\begin{equation}\label{eq:psi_in}
\psi^\mathrm{inc} = \sum_{j=\frac{1}{2}}^\infty a_j \left(j_{j+\frac{1}{2}}(kr)\mathcal{Y}_{jm_j}^{(-)}+\ii j_{j-\frac{1}{2}}(kr)\mathcal{Y}_{jm_j}^{(+)}\right),
\end{equation} 
which results in $a_j=-\ii^{j+\frac{1}{2}}\sqrt{4\pi(j+\frac{1}{2})}$ by using
the Rayleigh formula\cite{AbramowitzStegun} $\ee^{\ii kr\cos\theta}=\sum_{l=0}^{\infty}\ii^l(2l+1)j_l(kr)P_l(\cos\theta)$. Note that for such a spin-\textonehalf \space incoming plane wave, only the $m_j=\frac{1}{2}$ component contributes, hence the irrelevance of $m_j$ in the expansion. The full solution for an arbitrary incident direction is provided in Appendix~\ref{App:ArbitraryIncident}, where not only $m_j=\frac{1}{2}$ component contributes. Here we focus on the simplest $\ket{z\uparrow}$ case to reveal the essential physics. 
Then the total wave function outside the potential barrier $\psi^> = \psi^\mathrm{inc}+\psi^\mathrm{ref}$ is  
\begin{align}\label{eq:psi_out}
\psi^>
=\sum_{j=\frac{1}{2}}^\infty \sum_{\alpha=\pm1} \ii^{\frac{\alpha+1}{2}}(a_j j_{j-\frac{1}{2}\alpha}+b_j h_{j-\frac{1}{2}\alpha}^{(1)} )\mathcal{Y}_{jm_j}^{(\alpha)} 
\end{align}
where we use \eqref{eq:sphere_eigen_in} to expand $\psi^\mathrm{ref}$ with coefficients $b_j$.
Inside the barrier, we use \eqref{eq:sphere_eigen_out} to expand $\psi^<$ with coefficients $c_j$.
From the continuity of a wavefunction, $\psi^>(r=a)=\psi^<(r=a)$, we get the expansion coefficients $b_j=-u_j a_j,c_j=-v_j a_j$ with
\begin{align}\label{eq:b_j-c_j}
u_j&=\frac{j_{j+\frac{1}{2}}(ka) j_{j-\frac{1}{2}}(k'a)-j_{j-\frac{1}{2}}(ka)j_{j+\frac{1}{2}}(k'a)}
{h_{j+\frac{1}{2}}^{(1)}(ka)j_{j-\frac{1}{2}}(k'a)-h_{j-\frac{1}{2}}^{(1)}(ka)j_{j+\frac{1}{2}}(k'a)} \nonumber
\\
v_j&=\frac{j_{j+\frac{1}{2}}(ka) h_{j-\frac{1}{2}}^{(1)}(ka)-j_{j-\frac{1}{2}}(ka)h_{j+\frac{1}{2}}^{(1)}(ka)}
{h_{j+\frac{1}{2}}^{(1)}(ka)j_{j-\frac{1}{2}}(k'a)-h_{j-\frac{1}{2}}^{(1)}(ka)j_{j+\frac{1}{2}}(k'a)}.
\end{align}
The dimensionless quantities $ka,k'a$ completely determine the system. Note also that because of the independence on $k'$, the nominator of $v_j$ is actually always equal to a constant $\ii/(ka)^2$. We point out that assuming a vanishing wavefunction at the boundary $r=a$, i.e., setting all $c_j=0$, for an infinitely high barrier with $V_0\rightarrow\infty$, will lead to contradiction, which is fundamentally related to the Klein tunneling property of relativistic fermions\cite{Klein,Klein1}. 


It is helpful to proceed with the partial wave analysis, in which the outgoing part $\ee^{\ii kr}$ of the incident wave will attain a phase shift $\ee^{-\ii 2\delta}$\mycomment{\cite{Sakurai}}. As explained in Appendix~\ref{App:partial_wave}, we have the $j$-dependent phase shift factor
\begin{equation}\label{eq:phaseshift}
\ee^{-2\ii\delta_j}=1-2u_j.
\end{equation}
In Fig.~\ref{fig:sigma}(a), we show in logarithmic scale the phase shift $\delta_j$ varying with the energy of the incident electron. The phase shift approaches zero as we decrease $ka$ and the higher order ones are much smaller\mycomment{ than those of lower order}. Henceforth, we use $v=1, \hbar = 1$ in all the figures.

\section{Scattering cross section, electronic focusing, and resonance effect}\label{Sec:results}

\subsection{Scattering cross section}\label{Sec:cross_section}
\begin{figure*}[htbp]
\centering
\includegraphics[scale=0.8]{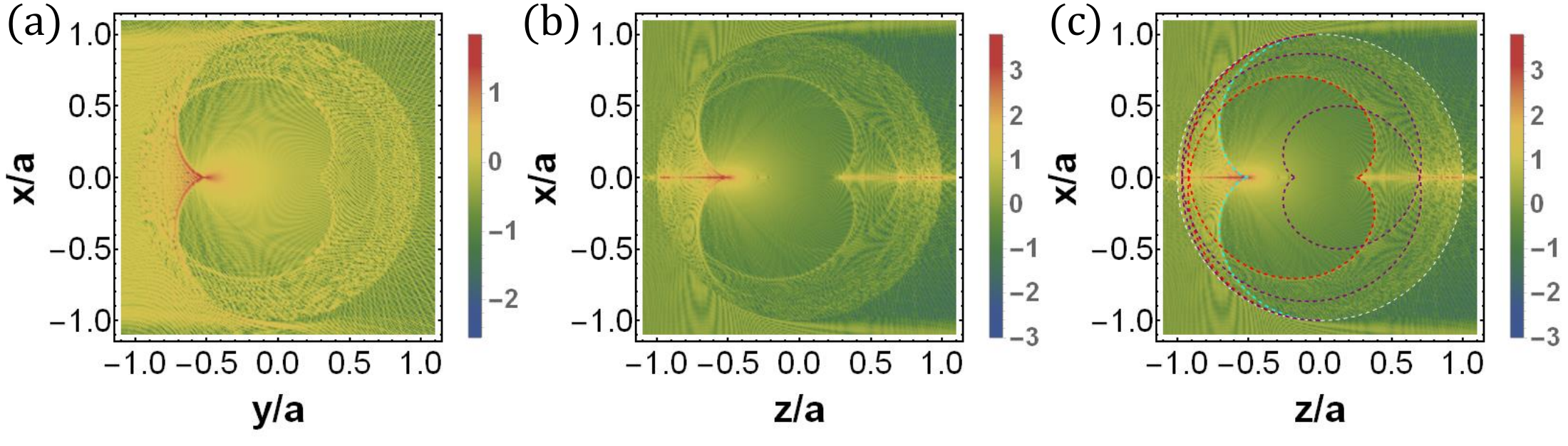}
\caption{(Color online) Electron probability density in the 2D $xy$-plane of graphene (a) and the $xz$-plane of 3D Weyl semimetal without (b) or with (c) the caustics curves predicted by geometric optics. The first (cyan), second (red), and third (purple) order caustics curves and the circular edge of the potential barrier (white) are shown in dashed lines. We set $ka=600,k'a=-600$, i.e., refractive index $n=-1$. The densities are plotted in a logarithmic scale. 
}\label{fig:caustics}
\end{figure*}
As detailed in Appendix~\ref{App:cross_section}, we start from the asymptotic form of the scattering amplitude $f(\theta,\phi)$ to study the scattering cross sections. Such information will be useful in designing electronic lens devices based on a Weyl semimetal. Firstly, the differential cross section is
\begin{equation}\label{eq:sigma_diff}
	\sigma(\theta,\phi)= \frac{2}{k^2(1+\cos\theta)}|\sum_{l=0}^\infty u_{l+\frac{1}{2}}(l+1)(P_{l+1}+P_l)|^2,
\end{equation}
where we introduce the Legendre polynomials $P_l$. Note that it does not really depend on $\phi$ and the far-field radial current density $j_r$ is by definition proportional to $\sigma(\theta,\phi)/r^2$, whose formula $j_r = v {\psi^\mathrm{ref}}^\dagger \vec{\sigma}\cdot\hat{r} \psi^\mathrm{ref}$ follows from the Heisenberg equation of motion. Therefore, the differential cross section provides the radiation characteristic. In Fig.~\ref{fig:sigma}(b,c), we plot two typical cases of $\sigma(\theta,\phi)$. Except the resonance effect discussed in Sec.~\ref{Sec:resonance}, the low-angular-momentum terms will dominate when the radius $a$ of the potential barrier is small, which is also justified by the partial wave analysis in Sec.~\ref{Sec:solution}. Retaining solely the leading order terms in $j$ and $ka$, we find in Appendix~\ref{App:cross_section} $\sigma(\theta,\phi)$ proportional to $1+\cos\theta$. This perfectly matches Fig.~\ref{fig:sigma}(b) where the backscattering is forbidden. When $a$ is large as shown in Fig.~\ref{fig:sigma}(c), many partial waves contribute to the scattering process and surprisingly form a very enhanced constructive interference pattern only for the forward scattering, which is reminiscent of the Poisson spot in wave optics. This is unusual since one would expect the presence of various reflected or refracted waves unparallel to the $z$-axis. Besides interference, these features should be partly related to the Klein tunneling phenomena of massless fermions, which makes the forward propagation easier. 


The total cross section and transport cross section are two important physical quantities. 
They are calculated in Appendix~\ref{App:cross_section} respectively as
\begin{align}
\sigma_\mathrm{tot} &= \int \dd\Omega \,\sigma(\theta,\phi) = \frac{8\pi}{k^2}\sum_{l=0}^\infty  (l+1)|u_{l+\frac{1}{2}}|^2\label{eq:sigma_tot}\\ 
\sigma_\mathrm{tr}&=\int \dd\Omega \,(1-\cos \theta)\sigma(\theta,\phi)\nonumber\\
&=\frac{8\pi}{k^2}\sum_{l=0}^\infty\frac{(l+1)(l+2)}{2l+3}|u_{l+\frac{1}{2}}-u_{l+\frac{3}{2}}|^2.\label{eq:sigma_tr}
\end{align}
And we have verified the optical theorem, $f(\theta=0)=\frac{k}{4\pi}\sigma_\mathrm{tot}$, which relates the forward scattering amplitude to the total cross section\cite{Sakurai}. 
Also, the electron mobility can be defined using the transport cross section $\mu = \frac{e}{\hbar k_F}\frac{1}{n_\mathrm{imp}\sigma_\mathrm{tr}}$ where $n_\mathrm{imp}$ is the impurity density and $\sigma_\mathrm{tr}$ depends on $v$. 
In experiments, the ratio between the two integrated cross sections corresponds to the ratio between the quantum life time and the transport life time, i.e., $\eta = \sigma_\mathrm{tot}/\sigma_\mathrm{tr}=\tau_\mathrm{q}/\tau_\mathrm{tr}$, where $\tau_\mathrm{q}$ can be determined from the damping rate of Shubnikov-de Haas oscillations while $\tau_\mathrm{tr}$ can be determined from the conductance without magnetic field in transport measurements\cite{GrapheneMobility,DiracMobility}. 
In general, a system with large $\eta$ implies the dominance of forward scatterings and small $\eta$ indicates that scatterings to other directions are present as well\cite{Mahan,GiulianiVignale2008}.

As seen in Fig.~\ref{fig:sigma}(d,e), when the barrier radius is relatively small, both $\sigma_\mathrm{tot}$ and $\sigma_\mathrm{tr}$ increase with the energy of incident electrons and so does the life time ratio $\eta$. When the incident energy or the radius is small, it becomes legitimate to retain the leading partial wave $u_\frac{1}{2}$ to find the ratio $\eta=3/2$, which is consistent with Fig.~\ref{fig:sigma}(e). This is unique to Weyl electrons since it is different from bith 2D graphene\cite{2DScatterCircular,Shytov2007} ($\eta=2$) and 3D massive fermions\cite{massive_cross_section,*atomicbook} ($\eta=1$). 
On the other hand, as the radius becomes larger, the behavior will be different. Shown in Fig.~\ref{fig:sigma}(f,g), $\sigma_\mathrm{tr}$ rapidly decreases with incident energy while $\sigma_\mathrm{tot}$ slightly increases after a nonmonotonic variation. The life time ratio $\eta$ still increases and can be very large. In accordance with the implications of $\eta$ aforementioned, this largeness of $\eta$ corroborates the sharp forward scattering peak in Fig.~\ref{fig:sigma}(c). Therefore, larger size of impurity potentials is possible to enhance of the mobility. This can be confirmed by measuring $\eta$ and will be the case when cluster impurity is the dominant type in a material. 


\subsection{Electronic caustics and focusing effect}\label{Sec:caustics}
\begin{figure*}[htbp]
\centering
\includegraphics[scale=0.5]{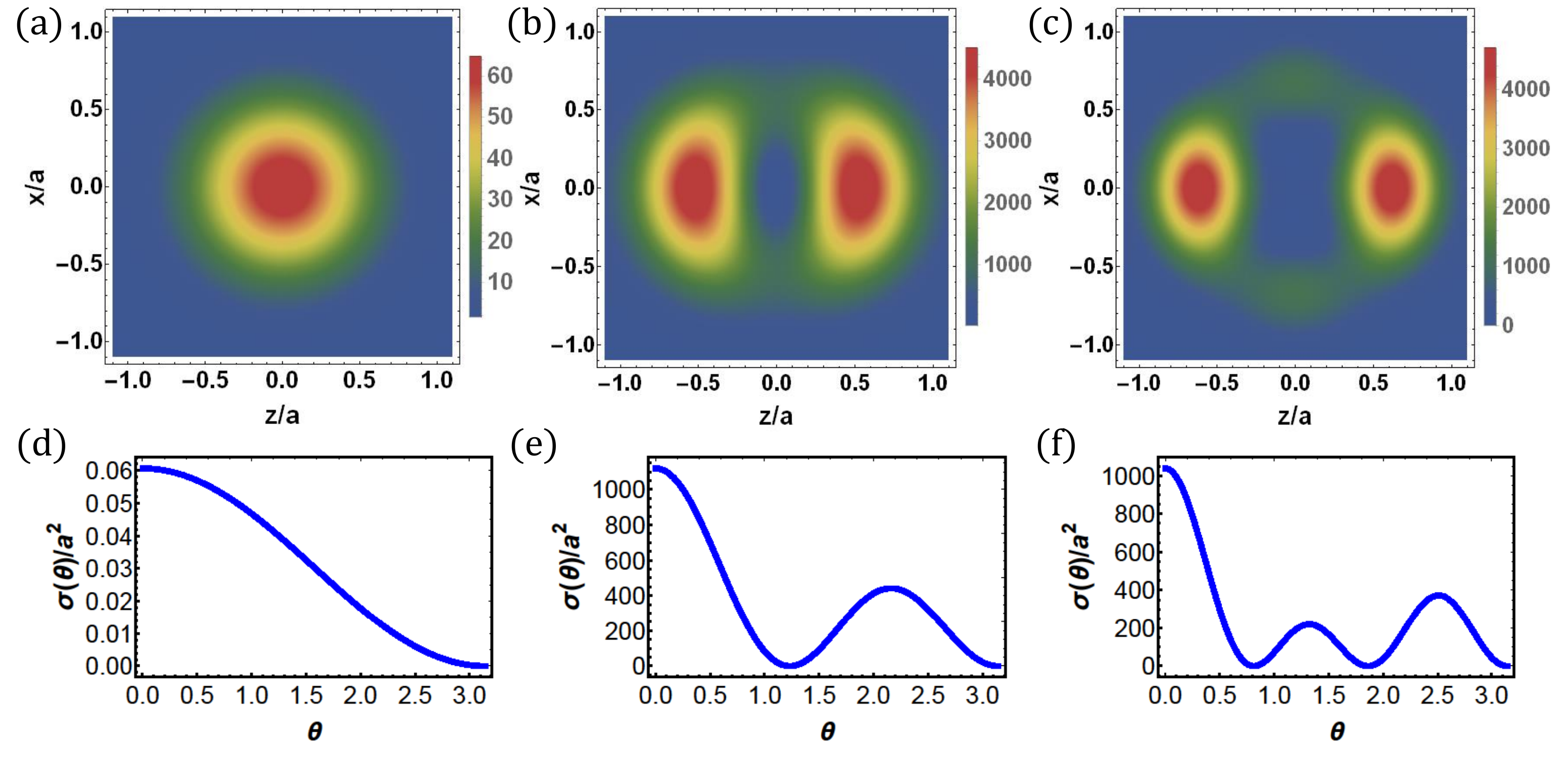}
\caption{(Color online) Resonances in a Weyl semimetal. (a,b,c): Near-field electron probability density. (d,e,f): Far-field differential cross section or radial current density. We set $a = 4, k = 0.02$. (a,d): $V_0 = 0.92$, no partial wave resonance, the main contribution comes from the $j=1/2$ partial wave. (b,e): $V_0 = 1.150\mycomment{1.14995373}$, the $j = 3/2$ partial wave resonates. Besides the global maximum at $\theta =0$, the current density has another local maximum. (c,f): $V_0=1.465\mycomment{1.46483326}$, the $j = 5/2$ partial wave resonates. The electron density keeps increasing and moving towards the poles. There are three maxima in the current density.}\label{fig:resonance}
\end{figure*}
Thanks to the linear dispersion, Weyl electrons share certain features with light. For example, it is predicted that the Goos-H\"anchen and Imbert-Fedorov shifts at the reflection interface in optics have their counterparts in 
 Weyl semimetals\cite{WeylShift1,WeylShift2,WeylShift3}. The making of a Veselago lens and the effects of tilted Weyl electrons with unidirectional barriers are also discussed\cite{Yesilyurt2016,VeselagoWeyl,tiltedWeyl}. The caustic is another aspect of interests in optics, which is the envelope line or surface of a family of reflected or refracted light rays and is often accompanied by cusp singularities\cite{OpticsBook,CausticBook}. Based on our exact solution, we can immediately show the existence of electronic caustics in a Weyl semimetal. The caustics can be calculated with Snell's law in geometric optics. The Weyl semimetal caustic surfaces can be obtained from the graphene case reported previously\cite{2DCaustics} by using the rotational invariance along the $z$-axis.

To observe the caustic phenomena as a classical limit, the electron wavelength must be much smaller than the radius of the potential barrier, i.e., $ka\gg 1$. Besides, large-angular-momentum components need to be included when using the exact solution in Sec.~\ref{Sec:solution} as required by Bohr's correspondence principle. Defining $n = k'/k$ as the refractive index, it is easy to have a negative one realized in a Weyl semimetal as long as the potential barrier $V_0>\hbar vk$, which is due to the linear dispersion. 
In Fig.~\ref{fig:caustics}, we plot, together with the graphene case, the electron probability density in the near field of the potential barrier in the $xz$-plane. We can see that the electron density exhibits a very clear pattern of caustics and cusps as calculated by Snell's law. Higher order caustics due to multiple refractions are less identifiable. 

In stark contrast to graphene without any special density profile on the symmetric axis, on long segments ($z<-0.5a$ and $z>0.3a$) 
along the spherical $z$-axis, we observe remarkably high electron density, which is in general two orders of magnitude larger than the peak value of the graphene case. This certainly differs from what one would expect from rotating the graphene density profile in terms of geometric optics. It is also robust against tuning $V_0$ or $n$, which can vary the pattern of caustics and cusps. We also note that on the reddish line segments, there exists several sub-red spots in addition to the obvious reddest cusp around $z=-0.5a$. Some sub-red spots can be partly related to higher order cusps or crossing points of caustics as exemplified by the ones around $z=-0.9a$ and $z=0.7a$, but not for, e.g., the absence of a red spot at the $z=-0.15a$ purple cusp. Since the above features cannot be understood by caustics or the equivalent geometric optics limit, we are led to regard them as quantum mechanical interference effects unique to the Weyl semimetal case. Correspondingly, in large regions, e.g., the one inside the red caustic curve and the one to the right of the white circular edge, the electron densities are actually much lower than those in the corresponding regions in graphene. This implies that the interference in fact focuses the electron probability cloud towards the two bright and reddish segments on the $z$-axis. It makes it a very intriguing possibility of application to electronic lens devices especially because the strong focusing is not only to a single point such as a cusp, but also along long line segments.

\subsection{Resonance in Weyl semimetal}\label{Sec:resonance}

When the energy of the incident electron and the radius of the potential are small, the system becomes more quantum mechanics dominant and the main contribution to the scattering comes from the lowest partial wave as mentioned in the last part of Sec.~\ref{Sec:solution}. However, we will show that this is not necessarily always the case by looking at both the near-field electron density profile and the far-field differential scattering cross section discussed in Sec.~\ref{Sec:cross_section}. 

The wavefunction probability density is virtually spherically symmetric and the differential cross section possesses a single maximum at $\theta=0$ as shown in Fig.~\ref{fig:resonance}(a,d), which is similar to Fig.~\ref{fig:sigma}(b). If we tune the incident energy or the potential, resonances of other partial waves can occur. In the coefficients \eqref{eq:b_j-c_j}, it is the imaginary part of the same denominator that is dramatically reduced at a certain $j$ by the resonant fine-tuning. Additionally, the nominator of $u_j$ is rapidly decreasing with $j$ while the nominator of $v_j$ is fixed. As a result, the electron density upsurges overall and reaches maximal values inside the potential. On the other hand, the angular dependence purely enters through the spin-\textonehalf \space angular wavefunctions
$\mathcal{Y}_{j=l\pm\frac{1}{2},m_j=\frac{1}{2}}^{(\pm)}$, which involve the spherical harmonics $Y_{l\pm1}^{m=0,1}$ as discussed in Sec.~\ref{Sec:solution}. Although the density or cross section generally have interferences between these wavefunctions, we can gain some insight from two properties of spherical harmonics\cite{ArfkenWeberHarris}. Firstly, the smaller $|m|$ usually means that larger amplitudes appear away from the equator and especially zonal ones ($m=0$) are peaked at the poles. Secondly, the nodal lines of wavefunction checker the latitudinal direction $l-|m|$ times. Indeed, as seen in Fig.~\ref{fig:resonance}(b,e,c,f), contrary to the spherically symmetric resonance and degenerate maxima in graphene\cite{2DMie1}, the electron density inside the potential is no longer spherically symmetric and tends to aggregate towards the south and north pole regions of the potential barrier while the far-field current density shows $j+\frac{1}{2}$ unequal peaks and nodes. 
These resonance phenomena indicate quasibound states formed near the potential barrier, which can be confirmed through two-terminal conductance measurements. It is possibly related to the integrability of the classical dynamics associated with closed paths of multiple oblique reflections inside the spherical region\cite{Integrable1,*Integrable2}. The polar profile can also be used as a distinguishing feature in experiments. As a whole, they may lead to interesting electronic lens applications.

\section{Summary}\label{Sec:summary}
In this paper, we provide the exact spherical eigenstates of the Weyl Hamiltonian and solve the scattering of a Weyl electron by a step potential barrier. Based on this solution, we compute the far-field current density and the total and transport scattering cross sections to discuss the related transport features at small or large radius of the potential barrier. A preferred forward scattering is found, which dominates at large energy or barrier radius. Using the wavefunction at the quasiclassical regime, although the analog of ray optical caustics is produced, a strong focusing effect on the high-symmetry axis due to quantum interference also shows up. For higher-angular-momentum resonances, we observe unequal directional far-field current peaks and a non-spherically symmetric profile of electron density whose polar shape gets enhanced with the resonance order. By showing the features unique to the Weyl electron in the important scenario of 3D spherical scattering, the present work can lead to more theoretical and experimental studies to exploit this new relativistic quasiparticle.


\section*{Acknowledgments}
M.L. was supported by NBRP (Nos.~2012CB821402 and 2015CB921102) and NSFC (Nos.~11534001, 11504008 and 11304280). X.-X.Z was supported by JSPS Grant-in-Aids for Scientific Research (Nos.~26103006 and 16J07545) and CREST (No.~JPMJCR16F1). 

\onecolumngrid
\appendix

\section{Solution for an arbitrary incident direction}\label{App:ArbitraryIncident}
If the incident direction $\vec{k}$ is not along the $z$-axis, the spin-momentum locking requires the spin to be in the state $\ket{\vec{k}\uparrow}=(\cos\frac{\theta_k}{2}\,,\sin\frac{\theta_k}{2}\ee^{\ii\phi_k})^T$, i.e., to align towards a general $\vec{k}=k\hat{k}=k(\sin\theta_k\cos\phi_k,\sin\theta_k\sin\phi_k,\cos\theta_k)$. Therefore, we have $m_j\neq\frac{1}{2}$ contributions. We can expand and compare the two sides of 
$
\ee^{\ii\vec{k}\cdot\vec{r}}\ket{\vec{k}\uparrow}
= \sum_{jm_j}a_{jm_j}(\varphi_{kjm_j}^{(-)}+\varphi_{kjm_j}^{(+)}).
$
After lengthy manipulations, 
we can obtain
\begin{equation}
a_{jm_j}=-\sqrt{4\pi}\,\ii^{j+\frac{1}{2}}\ee^{-\ii(m_j-\frac{1}{2})}\sqrt{\frac{(j-m_j)!}{(j+m_j)!}}\left[(j+1-m_j)\cos\frac{\theta_k}{2}P_{j+\frac{1}{2}}^{m_j-\frac{1}{2}}-\sin\frac{\theta_k}{2}P_{j+\frac{1}{2}}^{m_j+\frac{1}{2}} \right],
\end{equation}
where we introduce the associated Legendre polynomials.
We can verify the solution to be consistent with the $z$-axis incidence case when $\theta_k=0$. Since $P_{j+\frac{1}{2}}^{m_j-\frac{1}{2}}(1)\neq0$ only when $m_j=\frac{1}{2}$ and $P_l(1)=1$, we immediately find $a_{j,m_j=\frac{1}{2}}=-\sqrt{4\pi(j+\frac{1}{2})}\ii^{j+\frac{1}{2}}$ shown below \eqref{eq:psi_in}.

In analogy with the $z$-axis incidence case, we have the wavefunction outside the barrier
\begin{align}
\psi^> \mycomment{&= \ee^{\ii\vec{k}\cdot\vec{r}} \ket{\vec{k}\uparrow}
+\sum_{j=\frac{1}{2}}^\infty \sum_{m_j=-j}^j  b_{jm_j}(h_{j+\frac{1}{2}}^{(1)}(kr)\mathcal{Y}_{jm_j}^{(-)}+ \ii h_{j-\frac{1}{2}}^{(1)}(kr)\mathcal{Y}_{jm_j}^{(+)})\nonumber\\
&}=\sum_{j=\frac{1}{2}}^\infty\sum_{m_j=-j}^j (a_{jm_j} j_{j+\frac{1}{2}}+b_{jm_j} h_{j+\frac{1}{2}}^{(1)} )\mathcal{Y}_{jm_j}^{(-)} +  \ii(a_{jm_j} j_{j-\frac{1}{2}}+b_{jm_j} h_{j-\frac{1}{2}}^{(1)} )\mathcal{Y}_{jm_j}^{(+)}
\end{align}
and the wavefunction inside the barrier
\begin{equation}
\psi^<= \sum_{j=\frac{1}{2}}^\infty\sum_{m_j=-j}^j c_{jm_j} (j_{j+\frac{1}{2}}(k'r)\mathcal{Y}_{jm_j}^{(-)}+ \ii j_{j-\frac{1}{2}}(k'r)\mathcal{Y}_{jm_j}^{(+)}).
\end{equation}
From the continuity of a wavefunction, $\psi^>(r=a)=\psi^<(r=a)$,
we solve to find
\begin{align}\label{coefficients}
b_{jm_j}=-\frac{j_{j+\frac{1}{2}}(ka) j_{j-\frac{1}{2}}(k'a)-j_{j-\frac{1}{2}}(ka)j_{j+\frac{1}{2}}(k'a)}
{h_{j+\frac{1}{2}}^{(1)}(ka)j_{j-\frac{1}{2}}(k'a)-h_{j-\frac{1}{2}}^{(1)}(ka)j_{j+\frac{1}{2}}(k'a)} a_{jm_j},\quad\mycomment{\nonumber \\}
c_{jm_j}=-\frac{j_{j+\frac{1}{2}}(ka) h_{j-\frac{1}{2}}^{(1)}(ka)-j_{j-\frac{1}{2}}(ka)h_{j+\frac{1}{2}}^{(1)}(ka)}
{h_{j+\frac{1}{2}}^{(1)}(ka)j_{j-\frac{1}{2}}(k'a)-h_{j-\frac{1}{2}}^{(1)}(ka)j_{j+\frac{1}{2}}(k'a)} a_{jm_j}.
\end{align}
Again, when $ka\ll 1$, we need only keep $b_{\frac{1}{2},\pm\frac{1}{2}}$ terms.

\section{Partial wave analysis}\label{App:partial_wave}
In the partial wave expansion, \eqref{eq:psi_in} becomes 
\begin{equation}
\psi_\infty^>=\sum_{j=\frac{1}{2}}^{\infty}\frac{a_j}{2kr}\left\{ \left[(-\ii)^{j+\frac{3}{2}}\ee^{\ii (kr-2\delta_j)}+\ii^{j+\frac{3}{2}}\ee^{-\ii kr}\right]\mathcal{Y}_{jm_j}^{(-)}+\left[(-\ii)^{j+\frac{1}{2}}\ee^{\ii (kr-2\delta_j)}+\ii^{j+\frac{1}{2}}\ee^{-\ii kr}\right]\mathcal{Y}_{jm_j}^{(+)}\right\}
\end{equation}
while we have from \eqref{eq:psi_out}
\begin{align}\label{eq:out_asympt}
\psi_\infty^> =\sum_{j=\frac{1}{2}}^{\infty}\frac{a_j}{2kr}\left\{ \left[(1-2u_j)(-\ii)^{j+\frac{3}{2}}\ee^{\ii kr}+(\ii)^{j+\frac{3}{2}}\ee^{-\ii kr}\right]\mathcal{Y}_{jm_j}^{(-)}+\left[(1-2u_j)(-\ii)^{j+\frac{1}{2}}\ee^{\ii kr}+\ii^{j+\frac{1}{2}}\ee^{-\ii kr}\right]\mathcal{Y}_{jm_j}^{(+)}\right\},
\end{align}
where we use the asymptotic form of spherical Hankel functions\cite{AbramowitzStegun}, $h_l^{(1)}(x)\to \frac{1}{x}(-\ii)^{l+1} \ee^{\ii x}$, $h_l^{(2)}(x)\to \frac{1}{x}\ii^{l+1} \ee^{-\ii x}$.
Comparing them and using \eqref{eq:b_j-c_j}, we have \eqref{eq:phaseshift} or more explicitly,
\begin{equation}
\tan\delta_j = -\frac{j_{j+\frac{1}{2}}(ka)j_{j-\frac{1}{2}}(k'a)-j_{j-\frac{1}{2}}(ka)j_{j+\frac{1}{2}}(k'a)}{n_{j+\frac{1}{2}}(ka)j_{j-\frac{1}{2}}(k'a)-n_{j-\frac{1}{2}}(ka)j_{j+\frac{1}{2}}(k'a)}.
\end{equation}

\section{Scattering cross sections}\label{App:cross_section}
The scattering amplitude $f(\theta,\phi)$ 
can be read from the asymptotic form of $\psi^\mathrm{ref}$ that is included in \eqref{eq:out_asympt}
\begin{equation}\label{eq:scattering_amplitude}
f(\theta,\phi) = \sum_{j=\frac{1}{2}}^\infty \frac{b_j(-\ii)^{j+\frac{3}{2}}}{k}(\mathcal{Y}_{jm_j=1/2}^{(-)}-\mathcal{Y}_{jm_j=1/2}^{(+)})=\frac{-\ii}{k} \sum_{j=\frac{1}{2}}^\infty\sqrt{4\pi(j+\frac{1}{2}})u_j(\mathcal{Y}_{jm_j=1/2}^{(-)}-\mathcal{Y}_{jm_j=1/2}^{(+)}).
\end{equation}
The differential scattering cross section is
\begin{align}
\sigma(\theta,\phi)&=|f(\theta,\phi)|^2\mycomment{\nonumber
\\
&}=\frac{4\pi}{k^2}\sum_{j',j=\frac{1}{2}}^\infty \sqrt{(j'+\frac{1}{2})(j+\frac{1}{2})}u_{j'}^*u_j(\mathcal{Y}_{j'm_{j'}=1/2}^{(-)}-\mathcal{Y}_{j'm_{j'}=1/2}^{(+)}) (\mathcal{Y}_{jm_j=1/2}^{(-)}-\mathcal{Y}_{jm_j=1/2}^{(+)})\nonumber
\\
&= \frac{1}{k^2} \sum_{l',l=0}^\infty u_{l'+\frac{1}{2}}^*u_{l+\frac{1}{2}} \left[(l'+1)(l+1)(P_{l'+1}+P_{l'})(P_{l+1}+P_{l}) + (P_{l'+1}^1-P_{l'}^1)(P_{l+1}^1-P_{l}^1)\right]
\end{align}
Using $P_{l+1}^1-P_l^1=-\sqrt{\frac{1-x}{1+x}}(l+1)(P_l+P_{l+1})$ and henceforth occasionally denoting $\cos\theta$ by $x$, we can derive \eqref{eq:sigma_diff}
\begin{align}
\sigma(\theta,\phi)&= \frac{1}{k^2} \sum_{l',l=0}^\infty \frac{2}{1-x} u_{l'+\frac{1}{2}}^*u_{l+\frac{1}{2}} (P_{l'+1}^1-P_{l'}^1)(P_{l+1}^1-P_{l}^1)
=\frac{2}{k^2(1-x)}|\sum_{l=0}^\infty u_{l+\frac{1}{2}}(P_{l+1}^1-P_l^1)|^2\nonumber
\\
&=\frac{2}{k^2(1-x)}|\sum_{l=0}^\infty (u_{l+\frac{1}{2}}-u_{l+\frac{3}{2}})P_{l+1}^1|^2
=\frac{2}{k^2(1+x)}|\sum_{l=0}^\infty u_{l+\frac{1}{2}}(l+1)(P_{l+1}+P_l)|^2.
\end{align}
Using the orthogonality relations of (associated) Legendre polynomials\cite{AbramowitzStegun}
, we can obtain \eqref{eq:sigma_tot}. 
We can rewrite \eqref{eq:scattering_amplitude} with $\hat{b}_j=\frac{(-i)^{j+\frac{3}{2}} b_{j}}{\sqrt{4\pi(j+\frac{1}{2})}ka}$
\begin{equation}
f(\theta,\phi)=\hat{b}_{\frac{1}{2}}a\begin{pmatrix}
-(P_1^0+P_0^0)\\P_1^1 \ee^{\ii\phi}
\end{pmatrix}+
\sum_{j=3/2}^\infty \hat{b}_{j}a\begin{pmatrix} -(j+\frac{1}{2})(P_{j+\frac{1}{2}}^0+P_{j-\frac{1}{2}}^0)\\
(P_{j+\frac{1}{2}}^1-P_{j-\frac{1}{2}}^1)\ee^{\ii\phi}
\end{pmatrix}.
\end{equation}
If $ka\ll 1$, we can expand $\hat{b}_j$ to the leading nonvanishing order,
$\hat{b}_\frac{1}{2} = \left[\frac{1}{q_0a}-\cot(q_0a)\right]ka+O[(ka)^2]$ and 
$\hat{b}_\frac{3}{2} = \left[\frac{1}{3q_0a}-\frac{q_0a}{9[1-\cot(q_0a)]}\right](ka)^3+O[(ka)^4]$ where $q_0=V_0/(\hbar v)$.
Therefore, we need only keep $\hat{b}_\frac{1}{2}$ to see 
$f(\theta,\phi)=-\hat{b}_{\frac{1}{2}}a(\cos\theta+1,\sin\theta \ee^{\ii\phi})^T$
and
\begin{equation}
|f(\theta,\phi)|^2=2(\cos\theta+1)\left[(q_0a)^{-1}-\cot(q_0a)\right]^2k^2a^4.
\end{equation}
The transport cross section \eqref{eq:sigma_tr} is calculated as
\begin{align}
\sigma_\mathrm{tr}&=\frac{4\pi}{k^2}\int_{-1}^{1}dx |\sum_{l=0}^\infty (u_{l+\frac{1}{2}}-u_{l+\frac{3}{2}})P_{l+1}^1|^2\mycomment{\nonumber
\\
&}=\frac{8\pi}{k^2}\sum_{l=0}^\infty\frac{(l+1)(l+2)}{2l+3}|u_{l+\frac{1}{2}}-u_{l+\frac{3}{2}}|^2.
\end{align}

The optical theorem can be shown by using \eqref{eq:scattering_amplitude} and noting that $P_l(1)=1$ and $P_l^1(1)=0$
\begin{equation}
f(\theta=0,\phi)
=\frac{-\ii}{k} \sum_{j=\frac{1}{2}}^\infty\sqrt{4\pi(j+\frac{1}{2}})u_j(\mathcal{Y}_{jm_j=1/2}^{(-)}-\mathcal{Y}_{jm_j=1/2}^{(+)})
=\sum_{j=\frac{1}{2}}^\infty \frac{2\ii}{k}(j+\frac{1}{2})u_j
\end{equation}
Then from \eqref{eq:phaseshift} and \eqref{eq:sigma_tot}, we have
$\Im f(\theta=0)=\frac{2}{k}\sum_{j=\frac{1}{2}}^\infty (j+\frac{1}{2})\sin^2\delta_j =\frac{k}{4\pi}\sigma_\mathrm{tot}$.

\twocolumngrid

%

\bibliography{reference.bib}  

\end{document}